\documentstyle[12pt,epsf]{article}
 1
\def\citenum#1{{\def\@cite##1##2{##1}\cite{#1}}}
\def\citea#1{\@cite{#1}{}}

\def\Om{\Omega(s,b)}

\def\nex{\nu(s) e^{-\frac{b^{2}}{R^{2}(s)}}}

\def\beq{\begin{equation}}
\def\eeq{\end{equation}}
\def\bea{\begin{eqnarray}}
\def\eea{\end{eqnarray}}
\def\underarrow#1{\mathrel{\mathop{\longrightarrow}\limits_{#1}}}

\def\bbbz{{\mathchoice {\hbox{$\sf\textstyle Z\kern-0.4em Z$}}
{\hbox{$\sf\textstyle Z\kern-0.4em Z$}}
{\hbox{$\sf\scriptstyle Z\kern-0.3em Z$}}
{\hbox{$\sf\scriptscriptstyle Z\kern-0.2em Z$}}}}
\relax
\begin{document}
\begin{titlepage}
\noindent
 October 1993   \hfill  FERMILAB - PUB  - 93/000 - T \\TAUP 2066-93  \\
                       \\[9ex]
\begin{center}
{\Large \bf Diffractive Dissociation and      }   \\[1.4ex]
{\Large \bf Eikonalization in High Energy
         $ pp$ and     $p\bar p$ Collisions                  }   \\[11ex]

{\large E.\ Gotsman   *                         }    \\[1ex]
{\large E.M.\ Levin   ** $^{a)} $                          }    \\[1ex]
{\large U.\ Maor    *** $ ^{ b)} $              }    \\[1.5ex]
{*  School of Physics and Astronomy  }      \\
{Raymond and Beverly Sackler Faculty of Exact Sciences} \\
{Tel Aviv University, Tel Aviv 69978}  \\ [1.5ex]
 { **     Fermi National Accelerator Laboratory }  \\
{P.O. Box 500, Batavia, Illinois 60510} \\  [1.5ex]
{ *** Department of Physics } \\
{ University of Illinois at Urbana-Champaign, Urbana, Illinois 61801} \\
\footnote{ a) On leave from Petersburg Nuclear Physics Institute, Gatchina,
St. Petersburg}
\footnote{ b) On leave from the School of Physics and Astronomy,
 Tel Aviv University, Tel Aviv             }       \\ [5ex]
{\large \bf Abstract}
\end{center}
\begin{quotation}
We show that eikonal corrections imposed on diffraction dissociation
processes calculated in the triple Regge limit, produce a radical
change in the energy dependence of the predicted cross section.
The induced correction is shown to be in general
  agreement with the recent Tevatron experimental data.
\end{quotation}
\end{titlepage}
\par
    Over the past few years,
    phenomenological investigations of Pomeron exchange processes
 have been  almost exclusively confined to the
study of elastic scattering and total cross sections [1-5].
Recently published Tevatron data \cite{6,7} on single diffraction
dissociation (SDD), enables us to evaluate the compatability of
the parametrizations used to describe elastic and diffractive
scattering, and whether,
 it is necessary to include screening corrections,
to obtain a successful description of these processes.
\par A fundamental problem that that must be tackled when one
  attempts to make a comprehensive analysis of the published
high energy data on SDD [6-11],  is the fact
that there is no unique, agreed upon, experimental definition
of SDD. Experimental groups have used different, and not always
mutually consistent methods of extracting the desired data.
 In addition, it is difficult to compare the values that the
 different experimental groups  give for $ \sigma_{sd} $, as
in their evaluation of $ \frac{d \sigma_{sd}}{d M^{2} dt} $, they  have
used diverse intergration limits for t and $ M^{2} $ .
Futhermore, their treatment of the correlations observed between
$ M^{2} $ and t are entirely different.
 \par With the above limitations in mind, we present in this
communication,
a general study of SDD, which is compatible with the analysis of
elastic scattering, and at the same time reproduces all the important
features of the experimental data measured in SDD at high energies.

\par  Even though the Pomeron was introduced into high energy
 physics more than 30 years ago, its exact definition and detailed
substructure remain an enigma. In contrast to standard Regge
trajectories, the Pomeron has no particles on the time-like sector
of its trajectory. Nevertheless, it is required both phenomenologically
, to describe the forward hadron-hadron scattering data, and
theoretically to ensure that Regge theory is self consistent. Indeed,
in a Reggeon field calculus the Pomeron is described as a ladder
of Reggeons yielding \cite{12} $ \alpha(0) $ = 1. We will refer
to this as the "soft Pomeron" .

\par A number of different models have been proposed to account
for the rising hadron-hadron cross sections: \\
 1) Donnachie and Landshoff \cite{1} have advocated an
ad hoc approach in which the soft Pomeron amplitude keeps its
traditional form
 with $ \alpha(0) = 1 + \Delta  \simeq 1.08 $ .
  This simple model  reproduces the
qualitative features of the experimental data remarkably well.  \\
 2) Alternatively, one may perceive the Pomeron as a two gluon
exchange \cite{13}, or more generally as a gluon ladder.
Lipatov \cite{14} has shown that such a ladder, when calculated
 within the framework of perturbative QCD, receives its major
contribution from high $ p_{\perp} $ gluon exchanges. These give rise
to a series of poles in the complex j-plane above unity.
The summation of these poles yield the "hard Pomeron" with
 $ \Delta = \frac{12}{\pi} \alpha_{s}ln 2 $.
 Bjorken has suggested \cite{15} that the
generic Pomeron may actually manifest itself in both soft and hard
modes, each contributing in a different kinematical domain.
Models based on  a hybrid Pomeron
 are very successful in reproducing the data \cite{4,5}.  \\
  3) In the QCD inspired model \cite{2,3}, the
growth of the total cross section is associated with the greater
probability of semi-hard gluons  to interact with increasing
energy.
In this case, the need to describe the data over
 a wide energy range also
 requires a hybrid model \cite{3} consisting of a  soft
q-q background and  semi-hard q-g and g-g interactions.
\par All the above models of the Pomeron have a
intrinsic powerlike $ s^{\Delta} $
 rise of the total hadronic cross section. We note
 \cite{4,16} that the Pomeron amplitude proposed  in \cite{1}
 violates s-channel unitarity,
  just above the Tevatron energy range, for small {\it b}. In general,
we expect the unitarity bound to induce screening
 effects which saturate the growth of $ \sigma_{tot} $, making
$ \sigma_{tot} \leq ln^{2} s $, which is compatible with the Froissart
 bound. Technically, this is most easily achieved through eikonalization
\cite{17}, in which the amplitude discussed above serves as the lowest
order input to the eikonal expansion. Even though in the eikonal
model one only sums over elastic rescattering, ignoring diffraction
in the intermediate states, it has the advantage of being simple
to apply.
In addition, it introduces the natural scale of the screening
corrections, and allows one to explore different models of the Pomeron.
\par The main purpose of this letter is to examine the role played by
eikonalization in SDD. This is investigated
utilizing a simple Regge-like Pomeron \cite{1}. Extending the same
formalism to include an input Lipatov type Pomeron is straightforward.
As the presentely available diffractive data is not sufficiently refined
to enable one to discriminate between these models of the Pomeron, we
shall not discuss it in detail here.

\par The simplest way to write down the eikonal formulae is to consider
 the scattering process in impact parameter space. Our amplitude is
normalised so that
 \begin{equation}
   \frac{d \sigma}{dt} = \pi \vert f(s,t) \vert ^{2}
\end{equation}

 \begin{equation}
   \sigma_{tot} = 4 \pi Im f(s,0)
\end{equation}
 The scattering amplitude in {\it b}-space is defined as
 \begin{equation}
 a(s,b) = \frac{1}{2 \pi} \int d{\bf q}\;\; e^{-i{\bf q.b}}
  f(s,t)
\end{equation}
where $ t= - q^{2} $ .    \\
 In this representation
 \begin{equation}
 \sigma_{tot} = 2 \int d{\bf b}\;\; Im a(s,b)
\end{equation}
 \begin{equation}
 \sigma_{el} = \int d{\bf b}\;\; \vert a(s,b) \vert^{2}
\end{equation}
  s-channel unitarity when written in the diagonlised form
 implies
 \begin{equation}
2 Im a(s,b) = \vert a(s,b) \vert^{2} + G_{in}(s,b)
\end{equation}
where
 \begin{equation}
\sigma_{in} = \int d{\bf b}\;\; G_{in}(s,b)
\end{equation}
\par
We list below several assumptions  that we make regarding the
eikonal model:

1)  At high energy $ a(s,b) $ is assumed to be pure imaginary
 and can be reduced to the simple form
 \begin{equation}
a(s,b) = i ( 1 - e^{- \Omega(s,b)} )
\end{equation}
where $ \Omega(s,b) $ is a real function.
Analyticity
and crossing symmetry are easily restored to our oversimplified
 parametrization by substituting
$ s^{\alpha} \rightarrow s^{\alpha} e^{-i \pi \alpha /2}  $ .

 2) From eq. (6) we can express $ G_{in}(s,b)$
   as
 \begin{equation}
G_{in}(s,b) =1 - e^{-2 \Omega(s,b)}
\end{equation}
where $ e^{- 2 \Omega(s,b)} $ denotes the probability that no
inelastic interaction takes place at impact parameter {\it b}.

3) We write the t-channel Pomeron exchange  as
 \begin{equation}
 \Om =\nu(s) e^{-\frac{b^{2}}{R^{2}(s)}}
\end{equation}
In the simple Regge pole model  with a trajectory
   $ \alpha_{P}(t) = 1 + \Delta + \alpha^{\prime}t $ . We have
 \beq
\nu(s) = \frac{\sigma_{0}}{2 \pi R^{2}(s)}(\frac{s}{s_{0}})^{\Delta}
  \eeq
where
\beq
 R^{2}(s) = 4 [ R^{2}_{0} + \alpha^{\prime} ln \frac{s}{s_{0}} ]
\eeq
and  $ \sigma_{0} = \sigma(s_{0}) $ .
Agreement with the
 $ pp \; ({\bar p} p ) $ data is obtained  with
 $ R^{2}_{0} $ = 5.2 $ GeV^{-2} $ and $ \alpha^{\prime}$ = 0.25
$ GeV^{-2}$ . \\
  Eqs.(10-12) lead to simple expressions for the total and inelastic
cross sections with $ \sigma_{el} = \sigma_{tot} - \sigma_{inel} $ (see Fig.
2a).
 \beq
  \sigma_{tot} = 2 \pi R^{2}(s)[ ln \nu(s)
 + C - Ei(- \nu(s))]
  \underarrow {\nu \gg 1} 2 \pi R^{2}(s)[ ln \nu(s) + C ]
\eeq
\beq
\sigma_{in} =
  \pi R^{2}(s) [ln 2 \nu (s) + C - Ei(- 2 \nu (s) ) ]
  \underarrow { \nu \gg 1 } \pi R^{2}(s) [ ln 2 \nu(s) + C  ]
\eeq
where
  $ Ei(x) = \int^{x}_{- \infty} \frac{e^{t}}{t} dt $  ,
and  C  = 0.5773 is the Euler constant.
\par
 The standard approach to evaluate single diffractive dissociation
is through
the 3-body optical theorem \cite{18}
leading to the   PPP and PPR diagrams of interest
( see Fig. 1 ).
 The appropriate cross section is
 \beq
 M^{2} \frac{d \sigma_{sd}}{dM^{2}dt}
   = ( \frac{s}{M^{2}})^{2 \Delta + 2 \alpha^{\prime} t}
   [ G_{PPP}(t) (\frac{M^{2}}{s_{0}})^{ \Delta }
   + G_{PPR}(t) (\frac{M^{2}}{s_{0}})^{- \frac{1}{2}} ]
 \eeq
where all of the relevant couplings have been absorbed into
 $ G_{PPP}(t) $ or $ G_{PPR}(t) $. $ M^{2} $ denotes the mass
of the diffractive system, and for the Regge trajectory we have
taken $ \alpha_{R} (t) = \frac{1}{2} + t $ .

 Eq. (15) can be rewritten in the impact parameter representation
 \begin{eqnarray}
 \frac{M^{2}d \sigma_{sd}}{dM^{2}}
&=&  G_{PPP} \sigma^{2}_{0} (\frac{s}{M^{2}})^{2 \Delta}
        (\frac{M^{2}}{s_{0}})^{ \Delta}
\frac{1}{[ \pi {\bar R}^{2}_{1}(\frac{s}{M^{2}})]^{2}
   \pi {\bar R}^{2}_{1}(\frac{M^{2}}{s_{0}})}
   \int d{\bf b} d{\bf b^{\prime}}
 e^{- \frac{2(b-b^{\prime })^{2}}{{\bar R}^{2}_{1}(\frac{s}{M^{2}})}
 - \frac{b^{\prime 2 }}{{\bar R}^{2}_{1}(\frac{M^{2}}{s_{0}})}}
         \nonumber \\
  & + &
     G_{PPR} \sigma^{2}_{0} (\frac{s}{M^{2}})^{2 \Delta}
        (\frac{M^{2}}{s_{0}})^{- \frac{1}{2} }
\frac{1}{[ \pi {\bar R}^{2}_{1}(\frac{s}{M^{2})}]^{2}
   \pi {\bar R}^{2}_{2}(\frac{M^{2}}{s_{0}})}
   \int d{\bf b}d {\bf b^{\prime}}
 e^{- \frac{2(b-b^{\prime })^{2}}{{\bar R}^{2}_{1}(\frac{s}{M^{2}})}
 - \frac{b^{\prime 2 }}{{\bar R}^{2}_{2}(\frac{M^{2}}{s_{0}})}}
\nonumber    \\
 \end{eqnarray}
where
 \beq
    {\bar R^{2}_{i}}(\frac{s}{M^{2}}) = 2 R^{2}_{0i} + r^{2}_{0i}
      + 4 \alpha^{ \prime} ln(\frac{s}{M^{2}})
 \eeq
   $ r_{0i} \leq 1 GeV^{-2} $ denotes the radius of the triple
 vertex \cite{19} .
  $ {\bar R}^{2}_{1}( \frac{s}{M^{2}}) = 2 B_{sd} $ , where
$ B_{sd} $ denotes
 the slope of the SDD cross section .
Upon integrating eq. (16) we have
 \beq
 M^{2} \frac{d \sigma_{sd}}{dM^{2}} =
  \frac{\sigma^{2}_{0}}{ 2 \pi {\bar R}^{2}_{1}(\frac{s}{M^{2}})}
   (\frac{s}{M^{2}})^{2 \Delta }
[ G_{PPP}(\frac{M^{2}}{s_{0}})^{ \Delta} +
G_{PPR} (\frac{M^{2}}{s_{0}})^{ -\frac{1}{2}}  ]
 \eeq
\par We will now comment on consequences of the above result and its
relavence when compared to experimental data [6-11] :   \\
 1) We expect the forward SDD differential nuclear slope to be
 in the range $ \frac{1}{2} B_{el} < B_{sd} < B_{el} $, where
 $ B_{el} = 2 R^{2}(s) $ denotes the appropriate elastic scattering
slope. In general, $ B_{sd} $ is $ M^{2} $ dependent. An explicit
logarithmic dependence is implied by the definition of
$ {\bar R}_{i}(\frac{s}{M^{2}}) $ in eq.(17). We also note that due
to the different  $ M^{2} $ power dependences, the PPR contribution  is
concentrated at lower
 values of $ M^{2} $ than  the PPP. For energies in the
 ISR-Tevatron range, where $ ln^{2} s \geq R^{2}_{0} $ ,we expect
qualitatively, that $ B_{sd} \geq \frac{1}{2} B_{el} $
 with a very moderate
$ ln (\frac{s}{M^{2}}) $ dependence. This is in agrement with the data.
 We are unable to make a numerical fit due to strong correlations
 between $ M^{2} $ and t, observed at
 small values of $ M^{2} $ .
  We strongly urge that  measurements of
$ B_{sd} $ be made for higher values of the
  mass spectrum, say $ M^{2} \geq $
 16 $ GeV^{2} $.   \\
2) The $ M^{2} $ dependence of the SDD cross section is dominated by
\newline
$ [G_{PPP}(M^{2})^{-(1+ \Delta)} + G_{PPR}(M^{2})^{-(1.5+2 \Delta)}] $.
If we  express this dependence by $ ( M^{2})^{- \alpha_{eff}} $,
we expect $ (\alpha_{eff} - 1) > \Delta $
 and that $ (\alpha_{eff} $ - 1)
  approaches $ \Delta $ from above in the limit of very high s
 when the importance of the PPR term diminishes. This behaviour
is corroborated by the two recent studies \cite{6,7} of the
 $ M^{2} $ distribution at the Tevatron. In passing we note,
that the experiments at the FNAL \cite{8}
and ISR \cite{9} reported approximate
scaling, i.e. a $ (M^{2})^{- 1} $ behaviour. This is most probably
 due to the much narrower $ M^{2} $ interval investigated.
The approximation in which we only consider  the PPP + PPR terms
is obviously not sufficient to describe data at lower energies,
where lower lying trajectories are important \cite{20}.   \\
 3) Eq. (18) predicts a strong powerlike $ s^{2 \Delta} $
 dependence of the differential as well as the integrated SDD
cross section. This is a much stronger energy dependence than the
predicted $ s^{ \Delta} $ behaviour of $ \sigma_{tot} $, and  clearly
 not compatible with either theory or data.
  Indeed, the CDF data \cite{7} taken
at $ \sqrt{s} $ = 546 and 1800 GeV show only a moderate 20\%
 increase of the appropriate cross sections. This should be compared
 with an 80\% increase expected from a $ s^{2 \Delta} $ behaviour
 with $ \Delta $ = 0.125, as reported by CDF \cite{7}.
\par
Obviously, eq.(14) violates unitarity. Unitarity is restored,
in the eikonal model, by
 multiplying the integrand of eq. (16) by $ e^{-2 \Omega(s,b)} $
( see Fig. 2b).
 The resulting cross section is
 \begin{eqnarray}
  \frac{M^{2}d \sigma_{sd}}{dM^{2}} =
&\;&       G_{PPP} \sigma^{2}_{0} (\frac{s}{M^{2}})^{2 \Delta}
        (\frac{M^{2}}{s_{0}})^{ \Delta}
  \frac{1}{[ \pi {\bar R}^{2}_{1}(\frac{s}{M^{2}})]^{2}
    \pi {\bar R}^{2}_{1}(\frac{M^{2}}{s_{0}})} \cdot
  \nonumber \\
   &\;&    \int d{\bf b}d{\bf b^{\prime}}
    e^{- \nex} \cdot
  e^{- \frac{2(b-b^{\prime })^{2}}{{\bar R}^{2}_{1}(\frac{s}{M^{2}})}
   - \frac{b^{\prime 2}}{{\bar R}^{2}_{1}(\frac{M^{2}}{s_{0}})}}
\nonumber \\
&\;& + \;\; G_{PPR} \sigma^{2}_{0} (\frac{s}{M^{2}})^{2 \Delta}
        (\frac{M^{2}}{s_{0}})^{ -\frac{1}{2}}
  \frac{1}{[ \pi {\bar R}^{2}_{1}(\frac{s}{M^{2}})]^{2}
    \pi {\bar R}^{2}_{2}(\frac{M^{2}}{s_{0}})} \cdot
  \nonumber \\
  &\;&    \int d { \bf b}d {\bf b^{\prime}}
    e^{- \nex} \cdot
  e^{- \frac{2(b-b^{\prime })^{2}}{{\bar R}^{2}_{1}(\frac{s}{M^{2}})}
   - \frac{b^{\prime 2}}{{\bar R}^{2}_{2}(\frac{M^{2}}{s_{0}})}}
 \end{eqnarray}
where
$ \nu(s) $ is given by eq. (11) and $ R^{2}(s) $ by eq. (12).
After integration we have
  \newpage
 \begin{eqnarray}
  \frac{M^{2}d \sigma_{sd}}{dM^{2}} =
  \frac{\sigma^{2}_{0}}{ 2 \pi {\bar R}^{2}_{1}(\frac{s}{M^{2}})}
         (\frac{s}{M^{2}})^{2 \Delta} \cdot
 [ G_{PPP}(\frac{M^{2}}{s_{0}})^{ \Delta}
  a_{1} \frac{1}{(2 \nu (s))^{a_{1}}} \gamma (a_{1},2 \nu (s))
   \nonumber    \\
 + G_{PPR} (\frac{M^{2}}{s_{0}})^{ - \frac{1}{2}}
  a_{2} \frac{1}{(2 \nu (s))^{a_{2}}} \gamma (a_{2},2 \nu (s)) ]
 \end{eqnarray}
where
 \beq
a_{i} = \frac{2 R^{2}(s)}{ {\bar R}^{2}_{1}(\frac{s}{M^{2}}) +
     2 {\bar R}^{2}_{i}(\frac{M^{2}}{s_{0}})}
\eeq
and $ \gamma(a,2 \nu) $ denotes the incomplete Euler gamma function
$ \gamma(a,2 \nu) = \int^{2 \nu}_{0} z^{a - 1} e^{-z} dz $ .
\par
We list below the important consequences of the expression we obtained
in eq. (20).   \\
 1)
The {\it b}-space SDD amplitude, which is the integrand of eq. (19),
differs from the intrinsic integrand of eq. (16) by the corrective
multiplicative factor $ e^{ - 2 \Omega(s,b)} $. Whereas, the
unabsorbed {\it b}-space SDD amplitude is central and can be approximated
by a Gaussian centered at {\it b} = 0, the corrected amplitude has a dip
at {\it b} = 0, and its Gaussian approximation is centered at some
{\it b} = $ b_{0} \neq $ 0. This behaviour suggests that the generalized
unitarity condition \cite{20} is satisfied. This is consistent
with the general pattern expected of SDD b-space amplitudes after
screening has been included \cite{21}.   \\
 2) Our qualitative observation that $ B_{sd} \geq \frac{1}{2} $
$ B_{el} $ is unchanged.
 We expect the ratio
 $ \frac{B_{sd}}{B_{el}} $ to grow with energy, up to a limiting
 value of 1.  \\
 3) The dominant $ M^{2} $ dependence of
$ \frac{d \sigma_{sd}}{d M^{2}} $ is identical to that determined from
 eq.(18). We stress, that the two properties of the triple Regge
model, those concerning the t and $ M^{2} $ dependence, which are in
agreement with experiment, are essentially unchanged once the eikonal
correction is made to the original SDD amplitude. \\
 4) Eq. (20) exhibits a weak  s-dependence. This is
best seen if we examine our result in the high energy limit, where
 we have $ a_{i} \rightarrow  $ 2, and
$ \gamma[a_{i}, 2 \nu(s)] \rightarrow \Gamma(2) $ . Thus the factor
$ s^{2 \Delta} $ is compensated by $ [\frac{1}{\nu(s)}]^{a_{i}} $
 and eq. (20) reduces to
 \beq
 M^{2} \frac{d\sigma_{sd} }{dM^{2}} =
    \pi \Gamma(2) \sigma _{0}^{2}  R^{2}(s)
  [G_{PPP} (\frac{M^{2}}{s_{0}})^{- \Delta } +
  G_{PPR} (\frac{M^{2}}{s_{0}})^{- (\frac{1}{2} + 2 \Delta )}]
\eeq
Since $ \sigma_{sd} $ is not very sensitive to the high $ M^{2}$
integration limit,
 we find that $ \sigma_{sd} $ depends on s only through
$ R^{2}(s) $.
Our result indicated that the changes induced by eikonalization
on $ \sigma_{tot} $ and $ \sigma_{sd} $ are quite different.
For $ \sigma_{tot} $ the input $ s^{\Delta}$ power behaviour is
 modified to $ ln^{2}$s, the energy scale at which this change
becomes appreciable is  at $ \sqrt{s}  \approx $ 3 TeV \cite{4}.
 For $ \sigma_{sd} $ the input $ s^{2 \Delta} $ power behaviour
is modified to $ln$s, this occurs at an energy scale which is
considerably lower i.e. $ \sqrt{s}  \approx $ 300 GeV. In addition
 we expect that $ \frac{\sigma_{sd}}{\sigma_{tot}}
  \underarrow{s \rightarrow \infty } $ 0.
To test this theoretical prediction, we need to know $ \Delta $
and the ratio between $ G_{PPP} $ and $ G_{PPR} $. These two
parameters are obviously correlated.
 Donnachie and Landshoff \cite{1} suggest a global
$ \sigma_{tot} $ fit with $ \Delta $ = 0.08. This choice is compatible
with the CDF $ M^{2} $ distribution, if the PPR contributes 40\% of the
integrated  $ \sigma_{sd} $ at 546 GeV. The above value quoted for
$ \Delta $, which was suggested in \cite{1}, used the E710 measurement
 \cite{23} of $ \sigma_{tot} $ at $ \sqrt{s} $ = 1800 GeV. A recent
CDF measurement \cite{24} at the same energy has a considerably higher
value for $ \sigma_{tot} $ which is consistent with a value of
$ \Delta$ =  0.11. The corresponding PPR contribution to $ \sigma_{sd} $
 at 546 GeV is now reduced to 15\%. Irrespective of which value
of $ \Delta $ we use, we are not able
  to find an adequate overall fit to  SDD data  measured over the
entire energy range  [6-11].
 We feel that this is due to
 the following experimental and theoretical difficulties:  \\
1) As we noted previously, comparing $ \sigma_{sd} $ values obtained
by different experiments is not very instructive, due to the diverse
constraints and algorithims used by the different groups. \\
2) To minimize experimental uncertainties we consider the two CDF
measurements at $ \sqrt{s} $ = 546 and 1800 GeV, where they find
  \cite{7} R = $ \frac{ \sigma_{sd}(1800)}{ \sigma_{sd}(546)} $
= 1.20 $ \pm $ 0.06, with 1.4 $ \leq M^{2} \leq $ 0.15s. If we
 take $ \Delta $ = 0.08 we predict a ratio of $ R_{PPP} $ = 1.35
 for the PPP term, and for the PPR term $ R_{PPR} $ = 1.25. Assuming
the PPR contribution to account for 40\% of the SDD cross section at
 546 GeV, we have a theoretical prediction of R = 1.31. For
$ \Delta$ = 0.11, we obtain $ R_{PPP} $ = 1.35  and
$ R_{PPR} $ = 1.20. This gives us a prediction for R = 1.33, assuming
the PPR to account for 15\% of the SDD cross section at 546 GeV. \\
3) The CDF group start their $ M^{2} $ integration at
$ M^{2}_{min} $ = 1.4 Ge$V^{2}$, which is much too low for any
triple Regge analysis. To eliminate  the region of  low diffractive
 masses, we compare with the experimental ratio quoted by CDF
\cite{25} of R = 1.24 $ \pm $ 0.10
 obtained with $ M^{2}_{min} $ = 16 Ge$V^{2} $.
For $ \Delta$ = 0.08 we obtain R = 1.34, while for  $ \Delta $
= 0.11 we have R = 1.37.  \\
 4) Extrapolation of our model to ISR energies ( using values of the
parameters normalised to the CDF data) underestimates the
measured values of $ \sigma_{sd} $. This is not unexpected, as our
simple model with only PPP + PPR contributions is clearly not
sufficient at ISR energies, where a more detailed analysis
\cite{20} demonstrates the importance of lower lying trajectories
 at these energies. Examining SDD data over the whole energy range
[6-11], it appears that screening corrections become important
 at energies lower than that predicted by our eikonal model.
This is not surprizing, as in our treatment of eikonalization we
have only included elastic rescattering effects in the intermediate
states, while completely ignoring diffractive effects or so
called inelastic shadowing correction (see Fig. 2c)\cite{26}.
 Such corrections cannot  be considered to be small as the ratio
 $\sigma_{sd} / \sigma_{el}$ is of the order of $\frac{1}{2}$ at the Tevatron
energies. It means that dimensionless triple Pomeron vertex introduced in
eq.(19) is about $\frac{1}{8}$ and  diagrams of Fig. 2c should be taken into
account at the next stage of our approach.\\
5) In contrast to point 4) we expect the extrapolation of our
results to extremely high energies to be trustworthy. Integrating over
 1.4 $ GeV^{2} \leq M^{2} \leq $ 0.15 s, we predict  that
$ \sigma_{sd}$ = 13.3  and 13.9 mb
 at $ \sqrt{s} $ =16 and 40 TeV respectively,
 demonstrating the very weak
s dependence predicted by our model.
\par In conclusion, we wish to emphasis that our model does
reproduce the main features of SDD above 300 GeV, in particular the
 exceedingly moderate dependence of $ \sigma_{sd} $ on s. The model
which does not include lower lying Regge trajectories is
too simple to  successfully describe the SDD data at lower energies.

{\bf Acknowledgements:} We thank P. Giromini who discussed with
us the CDF experimental data {\cite{7} \cite{24} prior to publication and
for many helpful comments. We would also like to thank
L.Frankfurt for some provocative questions and constructive criticism.
 E.L. and U.M.   wish  to acknowledge the
kind hospitality and support of their collegues at the Fermilab
Theory Group  and the High Energy Group at the
  University of Illinois  at Urbana-Champaign.
\newpage
\vglue 0.5cm
\vglue 0.4cm

\newpage
\vglue 0.5cm
\section*{Figure Captions}
\vglue 0.4cm

{\bf Fig.1:}  SDD in the triple Regge approximation.

{\bf Fig.2:}

a) Screening corrections in the eikonal approximation to
 elastic scattering.

b) Screening corrections in the eikonal approximation to SDD.

c) Inelastic  shadowing (screeening) corrections to SDD.
\end{document}